\documentstyle[12pt]{article}
\def\beq{\begin{equation}}
\def\eeq{\end{equation}}
\def\bea{\begin{eqnarray}}
\def\eea{\end{eqnarray}}
\def\nn{\nonumber}
\def\ba{\begin{array}}
\def\ea{\end{array}}
\def\d{\partial}
\def\v{\vert}
\def\l{\langle}
\def\r{\rangle}

\begin{document}
\rightline{University of Tokyo Preprint}
\rightline{UTPH/99-20}
\smallskip
\baselineskip16pt
\smallskip
\begin{center}
{\large \bf \sf
       Exact spectrum and partition function  \\
   of $SU(m|n)$ supersymmetric Polychronakos model  } \\
\vspace{1.75 cm}

{\sf 
Bireswar Basu-Mallick$^1$\footnote{E-mail: 
biru@monet.phys.s.u-tokyo.ac.jp }, ~ Hideaki Ujino$^2$\footnote{E-mail: 
ujino@monet.phys.s.u-tokyo.ac.jp } ~
and ~Miki Wadati$^1$\footnote{E-mail: wadati@monet.phys.s.u-tokyo.ac.jp } }

\medskip \bigskip

{\em $^1$Department of Physics, Faculty of Science, University of Tokyo, \\
 Hongo 7-3-1, Bunkyo-ku, Tokyo 113-0033, Japan }

\smallskip \bigskip

{\em $^2$Gunma College of Technology, \\
         580 Toriba-machi, \\
         Maebashi-shi, Gunma 371-8530, Japan }

\end{center}

\vspace {2.25 cm}
\baselineskip=20pt
\noindent {\bf Abstract }

By using the fact that Polychronakos-like models can be obtained 
through the `freezing limit' of related spin Calogero models, we
calculate the exact spectrum as well as partition function of $SU(m|n)$ 
supersymmetric Polychronakos (SP) model. It turns out that, similar to the 
non-supersymmetric case, the spectrum of $SU(m|n)$ SP model is also equally 
spaced.  However, the degeneracy factors of corresponding energy levels 
crucially depend on the values of bosonic degrees of freedom ($m$) and 
fermionic degrees of freedom ($n$). As a result, the partition 
functions of SP models are expressed through some novel $q$-polynomials.  
Finally, by interchanging the bosonic and fermionic degrees of freedom, 
we obtain a duality relation among the partition functions of SP
models.

\vspace{.45 cm}

$q$-polynomials

\newpage

\baselineskip=22pt
\noindent \section {Introduction }
\renewcommand{\theequation}{1.{\arabic{equation}}}
\setcounter{equation}{0}

\medskip

Integrable one-dimensional models with long ranged interactions 
have recently attracted a lot of interest due to their close connection 
with diverse subjects like fractional statistics,
random matrix theory, level statistics for disordered systems, 
 Yangian algebra, q-polynomials etc. [1-20]. The 
$SU(M)$  Polychronakos spin chain [6-8] is a well known 
 example of such integrable models with the Hamiltonian given by
\beq
 H_{P} ~=~  \sum_{ 1 \leq i <j \leq N } \,
{ \left ( 1- \epsilon P_{ij} \right ) \over \left ( {\bar x}_i 
- {\bar x}_j \right)^2 }
\, ,
\label {a1}
\eeq
where 
 $\epsilon = 1 (-1)$ represents the ferromagnetic (anti-ferromagnetic)
case and $P_{ij}$ is the exchange operator interchanging the `spins' 
(taking $M$ possible values) of $i$-th and $j$-th lattice sites. 
Moreover the positions of corresponding lattice sites (${\bar x}_i$), 
which are inhomogeneously distributed on a line, are given by the zero 
points of the $N$-th order Hermite polynomial and may also be obtained as
a solution of the following set of equations:
\beq
x_i ~=~ \sum_{k \neq i}  \, { 2  \over \left( x_i - x_k \right)^3  }\, ,
\label {a2}
\eeq
where $i \in [1,2, \cdots ,N]$. Though the Polychronakos spin chain 
(\ref {a1}) does not enjoy translational invariance, one can find out
its exact spectrum as well as partition function  by considering the 
`freezing limit' of the Calogero Hamiltonian which possesses both 
spin and particle degrees of freedom [8]: 
\beq
H_C ~=~ {1\over 2} \, \sum_{i=1}^N \, 
\left ( \, - \, { \d^2 \over \d x_i^2 }  + \omega^2 x_i^2 \, \right ) ~+~
   \sum_{ 1 \leq i <j \leq N } \,
{ l \left ( l - \epsilon P_{ij} \right ) \over \left ( x_i 
- x_j \right)^2 } \, ,
\label {a3}
\eeq
$\omega $ and $l$ being some positive coupling constants. Thus, it is 
revealed that the Hamiltonian (\ref {a1}) generates an equidistant spectrum,
where the energy levels are highly degenerate in general. 
Such high degeneracy of energy levels can be explained through the `motif' 
picture  originating from the $Y(gl_M)$ Yangian symmetry 
of the Polychronakos spin chain (\ref {a1}) and Calogero model (\ref {a3})
[13,21].  Moreover, the partition function 
of the Hamiltonian (\ref {a1}) is found to be closely related [13,22] to the 
$SU(M)$ Rogers-Szeg${\ddot {\rm o}}$ (RS) polynomial, which have appeared 
earlier in different contexts like the theory of partitions [23] and 
the character formula for the Heisenberg XXX spin chain [24].

Now, for the purpose of obtaining a supersymmetric 
extension of the Polychronakos  spin chain, we consider a set 
of operators like $C_{i \alpha}^\dagger $ ($C_{i \alpha}$), which  
creates (annihilates) a particle of species $\alpha $ on the $i$-th 
lattice site. Such creation (annihilation) operators are defined to be 
bosonic if $\alpha  \in [1,2, \cdots , m ] $ and fermionic 
if $\alpha  \in [m+1, m+2,  \cdots , m+n  ] $ (where, according to 
our notation, $m+n=M$). Next, we focus our attention only to 
a subspace of the related Fock space, for which the total number of 
particles per site is always one:
\beq 
\sum_{\alpha =1 }^M \, C_{i \alpha}^\dagger C_{i \alpha} ~=~ 1 \, ,
\label {a4}
\eeq
for all $i$. On the above mentioned subspace, one can 
define a supersymmetric exchange operator as [18]
\beq 
{\hat P}^{(m|n)}_{ij} ~=~ \sum_{\alpha , \beta = 1}^M \, 
 C_{i \alpha}^\dagger C_{j \beta}^\dagger 
 C_{i \beta }C_{j \alpha } \,,
\label {a5}
\eeq
and show that this ${\hat P}^{(m|n)}_{ij}$ 
 yields a realisation of the permutation algebra given by
\beq
  {\cal P}_{ij}^2 ~=~ 1,
~~~{\cal P}_{ij}{\cal P}_{jl} ~=~ {\cal P}_{il}{\cal P}_{ij} ~=~
{\cal P}_{jl} {\cal P}_{il} ~, ~~~[ {\cal P}_{ij},
{\cal P}_{lm} ] ~=~ 0 \, ,
\label {a6}
\eeq
($ i, ~j,~l,~m $  being all distinct). 
This supersymmetric exchange operator (\ref {a5}) was used earlier 
for constructing the  $SU(m|n)$ supersymmetric Haldane-Shastry
(HS) model [18]. So, in analogy with the case of supersymmetric HS model,
 we may use the exchange operator (\ref {a5}) for constructing a 
Hamiltonian of the form 
\beq
 {\cal H}^{(m|n)}_{P} ~=~  \sum_{ 1 \leq i <j \leq N } \,
{ \left ( \, 1- {\hat P}^{(m|n)}_{ij} \, \right ) \over \left ( {\bar x}_i 
- {\bar x}_j \right)^2 }
\, .
\label {a7}
\eeq
It should be noted that, at the special case $m=M, \, n=0$, i.e. when 
all degrees of freedom are bosonic, 
${\hat P}^{(m|n)}_{ij}$ (\ref {a5}) becomes equivalent 
to the spin exchange operator $P_{ij}$ which appears in the Hamiltonian
(\ref {a1}).  Therefore, at this pure bosonic case, (\ref {a7}) would 
 reproduce the ferromagnetic Polychronakos spin chain (\ref {a1})
where $\epsilon = 1 $. Similarly, for the case
 $m=0, \, n=M$,  i.e. , when all degrees of freedom are fermionic,
(\ref {a7}) would reproduce the anti-ferromagnetic 
Polychronakos spin chain (\ref {a1}) 
where $\epsilon = - 1 $. So, when both bosonic and fermionic 
degrees of freedom are involved (i.e.,  when both $m$ and $n$ take 
nonzero values), we may say that $ {\cal H}^{(m|n)}_{P} $ 
(\ref {a7}) represents the Hamiltonian of $SU(m|n)$ supersymmetric 
Polychronakos (SP) model.

In this article, our aim is to study the spectrum and partition 
function of the above defined $SU(m|n)$ SP model.
To this end, in Sec.2, we introduce an appropriate 
extension of spin Calogero model (\ref {a3}), which would generate the 
 SP model (\ref {a7}) at the `freezing limit'.
In this section, we also demonstrate that the spectrum of 
such extended spin Calogero model is essentially the same with the spectrum 
of decoupled $SU(m|n)$ supersymmetric harmonic 
oscillators, where each oscillator has $m$ bosonic 
as well as $n$ fermionic spin degrees of freedom.
In this way we are able to show that, the equidistant 
spectrum of $SU(m|n)$ SP model (\ref {a7}) can be obtained by simply 
`modding out' the spectrum of $SU(m|n)$ harmonic oscillators 
through the spectrum of spinless bosonic harmonic oscillators. 
Subsequently, in Sec.3, we derive the partition function of SP model
(\ref {a7}) and interestingly observe that such partition functions can be
expressed through some novel $q$-polynomials. Finally, we obtain a duality 
relation between the partition functions of $SU(m|n)$ and $SU(n|m)$ 
SP models. Sec.4 is the concluding section.

\vspace{1cm}

\noindent \section { Spectra of $SU(m|n)$ SP model 
  and related spin Calogero model }
\renewcommand{\theequation}{2.{\arabic{equation}}}
\setcounter{equation}{0}

\medskip

For obtaining the spectrum of SP model, we wish to follow here the 
approach of Ref.8 and find out at first 
a suitable extension of spin Calogero model (\ref {a3}), 
which would reproduce the SP model (\ref {a7}) at the 
`freezing limit'. However, one immediate problem which arises at present 
is that a spin Calogero model like (\ref {a3}) is a first quantised 
system, while the SP model (\ref {a7}) is a second 
quantised system. So, for applying the above mentioned approach,
it is convenient to transform the second quantised SP
 model (\ref {a7}) to a first quantised spin system. 

To this end, we now consider some special cases of `anyon like' 
representations of permutation algebra (\ref {a6}), which were used 
earlier for constructing integrable extensions of $SU(M)$ Calogero-Sutherland 
model as well as HS spin chain [25-28]. Such special cases of `anyon like' 
representations are defined by their action on a spin state like 
 $\v \alpha_1 \alpha_2 \cdots  \alpha_N \r $ (with 
  $\alpha_i \in [1,2, \cdots , M] \, $) as
\beq
{\tilde P}^{(m|n)}_{ij} \,
 \v \alpha_1 \alpha_2 \cdots \alpha_i \cdots \alpha_j
\cdots \alpha_N \r ~=~ e^{ i \Phi^{(m|n)}
 (\alpha_i , \alpha_{i+1} , \cdots , \alpha_j ) } \, \v \alpha_1
\alpha_2 \cdots \alpha_j \cdots \alpha_i \cdots \alpha_N \r \,   , ~
\label {b1}
\eeq
where  
$ \Phi^{(m|n)} (\alpha_i , \alpha_{i+1} , \cdots , \alpha_j ) \, = \, 0$
if $\alpha_i , \, \alpha_j \, \in [1,2, \cdots , m ]$, 
$~\Phi^{(m|n)} (\alpha_i , \alpha_{i+1} , \cdots , \alpha_j ) \, = \, \pi $
if $\alpha_i , \, \alpha_j \, \in [ m+1,m+2, \cdots , m+n ]$, and 
$~ \Phi^{(m|n)} (\alpha_i , \alpha_{i+1} , \cdots , \alpha_j ) \, =  \,
\pi \, \sum_{\tau = m+1}^{m+n} \sum_{p=i+1}^{j-1} $    
$ \delta_{\tau , \alpha_p}$ if $\alpha_i \in [ 1,2, \cdots , m]$ and 
 $\alpha_j \in [ m+1,m+2, \cdots , m+n]$ or vice versa. It is clear that
${\tilde P}^{(M|0)}_{ij} $ reproduces the original spin exchange 
operator $P_{ij}$ and ${\tilde P}^{(0|M)}_{ij} $  reproduces $-P_{ij}$.
Next we notice that, due to the constraint (\ref {a4}),
 the Hilbert space associated with SP Hamiltonian 
(\ref {a7}) can be spanned through the following orthonormal basis vectors: 
$ C_{1 \alpha_1 }^\dagger C_{2 \alpha_2 }^\dagger \cdots 
 C_{N \alpha_N }^\dagger  \, \v 0 \r $, where $\v 0 \r$ is the 
vacuum state and $\alpha_i \in [1,2, \cdots , M]$.  
So, it is possible to define a one-to-one correspondence between 
the state vectors of the above mentioned Hilbert space and the state 
 vectors associated with a spin chain as
\beq 
\v \alpha_1 \alpha_2 \cdots  \alpha_N \r ~ \leftrightarrow ~
 C_{1 \alpha_1 }^\dagger C_{2 \alpha_2 }^\dagger \cdots 
 C_{N \alpha_N }^\dagger  \, \v 0 \r  \, .
\label {b2}
\eeq
However it should be noted that, the 
 matrix elements of ${\tilde P}^{(m|n)}_{ij}$ (\ref {b1}) 
and ${\hat P}^{(m|n)}_{ij}$ (\ref {a5}) are related as [28]
\beq 
\l \alpha'_1 \alpha'_2 \cdots  \alpha'_N \v  {\tilde P}_{ij} 
\v \alpha_1 \alpha_2 \cdots  \alpha_N \r  =  \l 0 \v 
C_{N\alpha'_N } C_{N-1, \alpha'_{N-1}} \cdots 
C_{1\alpha'_1 }  {\hat P}^{(m|n)}_{ij} 
 C_{1 \alpha_1 }^\dagger C_{2 \alpha_2 }^\dagger \cdots 
 C_{N \alpha_N }^\dagger   \v 0 \r   ,
\label {b3}
\eeq
where $\alpha_i$s and $\alpha'_i$s may be chosen in all possible ways.
Thus one finds that, the `anyon like' representation (\ref {b1}) is 
in fact equivalent to the supersymmetric realisation (\ref {a5}).
Consequently, the first quantised spin Hamiltonian given by 
\beq
  H^{(m|n)}_{P} ~=~  \sum_{ 1 \leq i <j \leq N } \,
{ \left ( \, 1- {\tilde P}^{(m|n)}_{ij} \, \right ) 
\over \left ( {\bar x}_i - {\bar x}_j \right)^2 }
\, ,
\label {b4}
\eeq
would also be completely equivalent to the second quantised SP model 
(\ref {a7}). In particular, the Hamiltonians (\ref {a7}) and (\ref {b4}) 
would share the same spectrum and their eigenfunctions can be related 
through the correspondence (\ref {b2}). 

Next, we define a spin Calogero model as
\beq
H^{(m|n)}_C ~=~ {1\over 2} \, \sum_{i=1}^N \, 
\left ( \, - \,  { \d^2 \over \d x_i^2 }  + \omega^2 x_i^2 \, \right ) ~+~
   \sum_{ 1 \leq i <j \leq N } \,
{ l \left ( \, l - {\tilde P}^{(m|n)}_{ij} \, \right ) \over 
\left ( x_i - x_j \right)^2 } \, ,
\label {b5}
\eeq
where ${\tilde P}^{(m|n)}_{ij}$ is the `anyon like' 
representation  (\ref {b1}). 
At the special case $m=M, \, n=0$ ($m=0, \, n=M$), 
the above Hamiltonian reproduces the $SU(M)$ spin Calogero model 
(\ref {a3}) with $\epsilon = 1$ ($\epsilon = - 1$). 
Notice that the Hamiltonian (\ref {b5}) can be rewritten as 
\beq
H^{(m|n)}_C ~=~ H_0 ~+~ l \,  H^{(m|n)}_1 \, , 
\label {b6}
\eeq
where $H_0$ is the Hamiltonian for spinless Calogero model:
\beq
H_0 ~=~ {1\over 2} \, \sum_{i=1}^N \, 
\left ( \,  -  \, { \d^2 \over \d x_i^2 }  + \omega^2 x_i^2 \, \right ) ~+~
   \sum_{ 1 \leq i <j \leq N } \, { l \left (  l - 1 \right ) \over 
\left ( x_i - x_j \right)^2 } \, ,
\label {b7}
\eeq
and $ H^{(m|n)}_1 $ is obtained from 
  $H^{(m|n)}_{P}$ (\ref {b4}) by replacing ${\bar x}_i$ and ${\bar x}_j$
with $x_i$ and $x_j$ respectively. Thus, the operator 
 $ H^{(m|n)}_1 $  possesses both spin and particle degrees of freedom.
However, in analogy with the 
 case of usual spin Calogero model (\ref {a3}) [8], we may now consider the 
`freezing limit' of extended spin Calogero model (\ref {b5}). Such 
`freezing limit' is obtained by setting 
 $l= \omega $ in the Hamiltonian (\ref {b5})
 and finally taking its  $\omega \rightarrow \infty $ limit. It is evident
that, at this `freezing limit', particle and spin degrees of freedom
 of Hamiltonian (\ref {b5}) would decouple and the operator
 $ H^{(m|n)}_1 $  would be transformed to a pure spin Hamiltonian
where the fixed values of $x_i$s are determined through the minima
of potential energy associated with
 spinless Calogero Hamiltonian (\ref {b7}).
Since the solution of eqn.(\ref {a2}) leads to the minima 
of such potential energy, the operator 
$ H^{(m|n)}_1 $  will exactly reproduce the SP Hamiltonian 
(\ref {b4}) at the `freezing limit'.  Consequently, the eigenfunctions 
of spin Calogero model (\ref {b5}) will
  factorise into eigenfunctions of spinless Calogero model (\ref {b7}) 
 containing only particle degrees of freedom 
and $SU(m|n)$ SP model (\ref {b4}) containing only spin degrees of freedom.
Moreover, at this `freezing limit',
the energy eigenvalues (denoted by $E_{p,s}(\omega )$, 
where the subscripts $p$ and $s$ represent the particle and spin degrees 
of freedom respectively) of the full system (\ref {b5}) can be expressed as
\beq 
 E_{p,s}(\omega ) ~=~ E_{p}(\omega ) ~+~ \omega \, E_{s} \, ,
\label {b8}
\eeq
where $ E_{p}(\omega ) $  and $E_{s} $ denote the energy eigenvalues of 
spinless Calogero model (\ref {b7}) and SP spin chain (\ref {b4})
respectively.

It is clear from eqn.(\ref {b8}) that, by `modding out' the spectrum of 
spin Calogero model (\ref {b5}) through the spectrum of 
spinless Calogero model (\ref {b7}), one can construct the spectrum
of SP model (\ref {b4}) or (\ref {a7}). So, for the purpose of obtaining 
the spectrum of SP model, it is essential to find out 
at first the spectrum of spin Calogero model (\ref {b5}). To this end,
we consider the spinless Calogero model of distinguishable particles,  
which is described by a Hamiltonian of the form 
\beq
{\cal H}_0 ~=~ {1\over 2} \, \sum_{i=1}^N \, 
\left ( \, -  \, { \d^2 \over \d x_i^2 }  + \omega^2 x_i^2 \, \right ) ~+~
   \sum_{ 1 \leq i <j \leq N } \, { l \left (  l - K_{ij} \right ) \over 
\left ( x_i - x_j \right)^2 } \, ,
\label {b9}
\eeq
where $K_{ij}$ is the coordinate exchange operator:
$K_{ij} \psi (\cdots , x_i , \cdots , x_j  , \cdots ) = 
\hfil \break
 \psi (\cdots , x_j , \cdots , x_i , \cdots ) $. By restricting the 
action of the above Hamiltonian on completely symmetric wave functions,
we may recover the spinless Calogero model (\ref {b7}) of 
indistinguishable particles. It has been recently found that,
by using some similarity transformations, one can completely decouple 
all particle degrees of freedom of Calogero Hamiltonians (\ref {b7}) 
as well as (\ref {b9}) [29-33].
Such similarity transformations naturally lead to a very efficient method 
of calculating the eigenfunctions of spinless Calogero models.
In particular, the spectrum and eigenfunctions of 
${\cal H}_0 $ (\ref {b9}) can be obtained through a similarity 
transformation which maps this interacting 
 Hamiltonian  to a system of decoupled harmonic oscillators like 
\beq
H_{free}~=~ \omega \, \sum_{k=1}^N \, a_k^\dagger a_k \, , 
\label {b10}
\eeq
where 
$ a_k^\dagger = { 1\over \sqrt {2 \omega } } 
\left ( \omega x_k - {\d \over \d x_k } \right )$ and 
$ a_k = { 1\over \sqrt {2 \omega } } 
\left ( \omega x_k + {\d \over \d x_k } \right )$. The above 
mentioned similarity transformation is explicitly given by [32,33]
\beq
 {\cal S}^{-1} \, \left( \, {\cal H}_0 - E_g \, \right) \, {\cal S}
  ~=~ H_{free}  \, ,
\label {b11}
\eeq
where $E_g  \, = \, {1\over 2} N \omega  \, 
\left (\, Nl + ( 1-l) \, \right)$ and 
\beq
 {\cal S} ~=~\phi_g \, e^{- { 1 \over 4 \omega } {\cal O}_L } \,
         e^{ { 1 \over 4 \omega } \nabla^2 }
  \, e^{ {1 \over 2} \omega X^2  } \, ,
\label {b12}
\eeq
with 
\bea 
&&X^2  \, = \,  \sum_{j=1}^N \, x_j^2 \, , ~~~
 \nabla^2  \, = \, \sum_{j=1}^N \, { \d^2 \over \d x_j^2 } \, , ~~~
\phi_g \, = \, \prod_{1\leq j < k \leq N }  \, \v x_j - x_k \v^l \,
\exp \left (  \, - {\omega \over 2} \sum_{j=1}^N x_j^2  \, \right ) \, , 
\nn \\
&&{\cal O}_L ~=~ \sum_{j=1}^N \, { \d^2 \over \d x_j^2 } \, + \, l \,
\sum_{ j\neq k} \, \left \{ \,  {1\over  (x_j - x_k ) } 
\left (   { \d \over \d x_j } -{ \d \over \d x_k }   \right ) \, + \, 
{  { K_{jk} - 1  } \over  { (x_j - x_k )^2 }  } \, \right \} \,  .\nn
\eea
The operator $ {\cal O}_L $ is called the Lassalle operator.
As is well known, the eigenfunctions of decoupled oscillators 
(\ref {b10}) can be written in the form:
$ \v n_1 n_2 \cdots n_N \r  \, = \,  \prod_{j=1}^N \,  
 \left ( a_j^\dagger \right )^{n_j} \, \v 0 \r , $ where 
 $\v 0 \r $ is the corresponding vacuum state.  
The coordinate representation of such eigenfunctions,
having eigenvalues $\omega \sum_{j=1}^N n_j $, is given by
\beq 
\psi_{n_1 , n_2 , \cdots , n_N } (x_1 , x_2, \cdots , x_N ) ~=~
e^{ - { 1\over 2}\omega X^2 } \, \prod_{j=1}^N  \, H_{n_j}(x_j) \, ,
\label {b13}
\eeq
 where $H_{n_j}(x_j)$ is the Hermite polynomial of order $n_j$.
It should be noted that, the above eigenfunctions do not generally obey any 
symmetry property under the exchange of coordinates and, therefore, 
they are free from any statistics. 
Due to the similarity transformation (\ref {b11}), the eigenfunctions 
of spinless Calogero model (\ref {b9}) may be obtained through the 
eigenfunctions (\ref {b13}) of free oscillators as 
\beq
\chi_{n_1 , n_2 , \cdots , n_N } (x_1 , x_2, \cdots , x_N ) ~=~
{\cal S} \, \left ( \, 
e^{ - { 1\over 2}\omega X^2 } \, \prod_{j=1}^N  \, H_{n_j}(x_j) 
\, \right ) \,  ,
\label {b14}
\eeq
with 
$E_{ \{ n_i \}  } ~=~ E_g + \omega \sum_{j=1}^N n_j $
representing the corresponding eigenvalues.
So, apart from a constant energy shift ($E_g$) 
for all levels, the spectrum of 
spinless Calogero model (\ref {b9}) is exactly the same 
as the spectrum of decoupled Hamiltonian (\ref {b10}) containing 
distinguishable particles [32,33].  A comment might be in order.
The exponentiation of the Lassalle operator
$e^{-\frac{1}{4\omega}{\cal O}_{L}}$ yields essential singularities at
$x_{i}=x_{j}$, $i,j=1,2,\cdots,N$, when it operates on general multivariable
functions. However, such essential singularities are not generated when the
exponentiation of the Lassalle operator operates on multivariable
polynomials. Our using of the Lassalle operator in the expression
(\ref {b14}) is such a safe one. Details on the property of the Lassalle 
operators are given in refs. 32, 33.

In the following, we want to show that the eigenfunctions of 
spin Calogero model (\ref {b5}) can also be obtained through the 
eigenfunctions of decoupled harmonic oscillators, provided they
possess nondynamical spin degrees of freedom and 
obey some definite statistics. To this end, we introduce a projection 
operator $\Lambda_N^{(m|n)}$ which satisfies the relations like 
\beq
 K_{ij} {\tilde P}_{ij}^{(m|n)} \Lambda_N^{(m|n)} ~=~
 \Lambda_N^{(m|n)} K_{ij} {\tilde P}_{ij}^{(m|n)} ~=~
 \Lambda_N^{(m|n)}  \, ,
\label {b15}
\eeq
where $i,~ j \, \in \, [1,2, \cdots , N]$. Such a projector 
can be formally written through `transposition' operator 
 $\tau_{ij}^{(m|n)}$  ($= {\tilde P}_{ij}^{(m|n)} K_{ij}$) as 
\beq
\Lambda_N^{(m|n)} ~=~ \sum_p \, \sum_{\{i_k , j_k \}} 
 \, {\tau }_{i_1j_1}^{(m|n)} {\tau }_{i_2j_2}^{(m|n)}
 \cdots  {\tau }_{i_pj_p}^{(m|n)} \, ,
\label {b16}
\eeq
where the series of transposition
 $ {\tau }_{i_1j_1}^{(m|n)} {\tau }_{i_2j_2}^{(m|n)}
 \cdots  {\tau }_{i_pj_p}^{(m|n)} $ represents 
an element of the permutation group ($P_N$) associated with $N$ objects and,
 due to the summations on $p$ and  $\{i_k , j_k \} $,
each element of $P_N$ would appear only once in 
the r.h.s. of the above equation. For example, 
$\Lambda_2^{(m|n)}$ and $\Lambda_3^{(m|n)} $ are given by:
$ \Lambda_2^{(m|n)} ~=~ 1 + {\tau }_{12}^{(m|n)} \, , ~$  
$ \Lambda_3^{(m|n)} ~=~ 1 + {\tau }_{12}^{(m|n)} + 
  {\tau }_{23}^{(m|n)} + {\tau }_{13}^{(m|n)} + 
  {\tau }_{23}^{(m|n)} {\tau }_{12}^{(m|n)} + 
  {\tau }_{12}^{(m|n)} {\tau }_{23}^{(m|n)} \, . $
It should be noted that at the special case $m=M,~n=0$ ($m=0,~n=M$),
the projector $\Lambda_N^{(m|n)}$ (\ref {b16})  
completely symmetrises (antisymmetrises) any wave function 
 under simultaneous interchange of particle as well as
 spin degrees of freedom, and thus 
 projects the wave function to bosonic (fermionic) subspace. 
Next, we multiply the eigenfunction 
(\ref {b14}) of Calogero Hamiltonian (\ref {b9}) by an arbitrary spin state 
$ \v \alpha_1 \alpha_2 \cdots \alpha_N \r $ and subsequently apply 
the projector (\ref {b16}) for obtaining a wave function like 
\beq
 \Psi^{  \alpha_1 , \, \alpha_2 , \cdots , \, \alpha_N }_{ 
n_1 , \, n_2 , \cdots , \, n_N } (x ;\gamma ) ~=~ 
\l \gamma_1 \gamma_2 \cdots \gamma_N \v \,  \Lambda_N^{(m|n)} \, 
\left \{  \, {\cal S} \, \left ( \, e^{ - { 1\over 2}\omega X^2 } \, 
\prod_{j=1}^N  \, H_{n_j}(x_j) \, \right ) \, 
 \v \alpha_1 \alpha_2 \cdots \alpha_N \r \,  \right \} \, ,
\label {b17}
\eeq
where $x \equiv x_1 , x_2, \cdots , x_N $ and $\gamma \equiv 
 \gamma_1 , \gamma_2 ,  \cdots , \gamma_N $. Since the operator
$ {\cal H}_0 $ (\ref {b9}) commutes with $K_{ij}$ as well as 
$ \Lambda_N^{(m|n)}$ (\ref {b16}), the expression (\ref {b17}) again
 gives an eigenfunction of the Calogero Hamiltonian (\ref {b9}),
where each particle possesses some nondynamical spin degrees of freedom. 
Consequently,  with the help of relation (\ref {b15}) which 
allows one to `replace' $K_{ij}$ by ${\tilde P}_{ij}^{(m|n)}$,
one can show that 
 $  \Psi^{  \alpha_1 , \, \alpha_2 , \cdots , \, \alpha_N }_{ 
n_1 , \, n_2 , \cdots , \, n_N } (x;\gamma )$  (\ref {b17}) also gives an 
eigenfunction of spin Calogero model (\ref {b5}) with eigenvalue
$E_{ \{ n_i \} , \{ \alpha_i \} } ~=~ E_g + \omega \sum_{j=1}^N n_j $.

Thus we interestingly find that, 
eqn.(\ref {b17}) produces all eigenfunctions of 
spin Calogero model (\ref {b5}) through the known eigenfunctions of decoupled
harmonic oscillators (\ref {b10}). However it is important to notice that,
due to the existence of projector $ \Lambda_N^{(m|n)} $ in eqn.(\ref {b17}),
a definite correlation is now imposed among the 
 eigenfunctions of decoupled harmonic oscillators.
To demonstrate this point in an explicit way, we observe 
first that the operator ${\cal S}$ (\ref {b12}) 
commutes with $K_{ij}$ and the projector $\Lambda_N^{(m|n)}$ 
(\ref {b16}). By using such commutation relations, eqn.(\ref {b17}) 
can be rewritten as 
\beq
 \Psi^{  \alpha_1 , \, \alpha_2 , \cdots , \, \alpha_N }_{ 
n_1 , \, n_2 , \cdots , \, n_N } (x; \gamma) ~=~ {\cal S} \, 
 {\tilde \Psi}^{  \alpha_1 , \, \alpha_2 , \cdots , \, \alpha_N }_{ 
n_1 , \, n_2 , \cdots , \, n_N }(x; \gamma ) \, , 
\label {b18}
\eeq
where
\beq
{\tilde \Psi}^{  \alpha_1 , \, \alpha_2 , \cdots , \, \alpha_N }_{ 
n_1 , \, n_2 , \cdots , \, n_N } (x; \gamma ) ~=~ 
\l \gamma_1 \gamma_2 \cdots \gamma_N \v \,  \Lambda_N^{(m|n)} \, 
\left \{ \, \left ( \, 
e^{ - { 1\over 2}\omega X^2 } \,
\prod_{j=1}^N  \, H_{n_j}(x_j) \, \right ) \, 
 \v \alpha_1 \alpha_2 \cdots \alpha_N \r \,  \right \} \, .
\label {b19}
\eeq
This $ {\tilde \Psi}^{  \alpha_1 , \, \alpha_2 , \cdots , \, \alpha_N }_{ 
n_1 , \, n_2 , \cdots , \, n_N } (x; \gamma ) $  represents
 a correlated eigenfunction (with eigenvalue $ \omega \sum_{j=1}^N n_j $)
 for decoupled harmonic oscillators (\ref {b10}),  where each 
oscillator carries  $M$ number of nondynamical spin degrees of freedom. 
To determine the precise nature of the above mentioned correlation, 
  we use the relation (\ref {b15}) and replace $\Lambda_N^{(m|n)}$ by 
 $ \Lambda_N^{(m|n)} K_{ij} {\tilde P}_{ij}^{(m|n)} $ in the r.h.s.
of eqn.(\ref {b19}). Finally, by acting 
 $ {\tilde P}_{ij}^{(m|n)} $ on 
 $ \v \alpha_1 \alpha_2 \cdots \alpha_N \r $ and 
$K_{ij}$ on $  e^{ - { 1\over 2}\omega X^2 } 
\prod_{j=1}^N  H_{n_j}(x_j) \,  $, we find that
  the eigenfunction (\ref {b19}) of free oscillators must satisfy 
the following symmetry condition under simultaneous interchange of related 
particle and spin quantum numbers:
\beq
{\tilde \Psi}^{\alpha_1 ,  \cdots , \, \alpha_i , \cdots , \, \alpha_j ,
\cdots , \, \alpha_N }_{ n_1 , \cdots , \, n_i , \cdots , \, n_j , \cdots , 
 \, n_N } (x; \gamma ) ~=~ e^{ i \Phi^{(m|n)}
(\alpha_i , \alpha_{i+1} , \cdots , \alpha_j ) } \, 
 {\tilde \Psi}^{\alpha_1 ,   \cdots , \, \alpha_j , \cdots , \, \alpha_i ,
\cdots , \, \alpha_N }_{ n_1 , \cdots , \, n_j , \cdots , \,
 n_i , \cdots , \, n_N }  
(x; \gamma ) \, ,
\label {b20}
\eeq
$ \Phi^{(m|n)} (\alpha_i , \alpha_{i+1} , \cdots , \alpha_j ) $ being the 
same phase factor which appeared in eqn.(\ref {b1}). It is clear
from eqn.(\ref {b20}) that, for the case
 $\alpha_i , \, \alpha_j \in [ 1,2, \cdots , m]$,  the eigenfunction 
$ {\tilde \Psi}^{\alpha_1 ,  \cdots , \, \alpha_i , \cdots , 
 \, \alpha_j , \cdots , \, \alpha_N }_{ n_1 , \cdots , \,
 n_i , \cdots , \,  n_j , \cdots , \,
n_N } (x; \gamma ) $
 remains completely unchanged under simultaneous 
interchange of particle and spin quantum numbers. Thus $\alpha_i $
may be treated as a `bosonic' quantum number, if it takes any value 
ranging from $1$ to $m$.
Next, by using eqn.(\ref {b20}) for the case $\alpha_i , \, \alpha_j 
\in [ m+1, m+2, \cdots , m+n]$, it is easy to see that 
$ {\tilde \Psi}^{\alpha_1 ,  \cdots , \, \alpha_i , \cdots , 
\, \alpha_j , \cdots , \, \alpha_N }_{ n_1 , \cdots , 
\, n_i , \cdots , \, n_j , \cdots ,  \,
n_N } (x; \gamma ) $ would pick up a minus sign under simultaneous 
interchange of particle and spin quantum numbers. 
Therefore, the eigenfunction
 $ {\tilde \Psi}^{\alpha_1 ,  \cdots , \, \alpha_i , \cdots , 
\, \alpha_j , \cdots , 
\, \alpha_N }_{ n_1 , \cdots , \, n_i , \cdots , \, n_j , \cdots , \,
n_N } (x; \gamma ) $ must be trivial if we choose $n_i=n_j$ and 
 $\alpha_i  = \alpha_j  \in [ m+1, m+2, \cdots , m+n]$. 
Thus $\alpha_i $ may be treated as a `fermionic' quantum number, 
if it takes any value ranging from $m+1$ to $m+n$.
  Notice that if we simultaneously
interchange a `bosonic' quantum number $\alpha_i$ 
with a `fermionic' quantum number $\alpha_j$ and $n_i$ with $n_j$,
then the eigenfunctions satisfying transformation relation 
 (\ref {b20}) would either remain invariant or 
pick up a minus sign depending on whether even or odd
number of fermionic spin quantum numbers are present in the configuration:
$ \alpha_{i+1}  \alpha_{i+2} \cdots \alpha_{j-1}$.
It is obvious that, at the special case $m=M, \, n=0$ ($m=0, \, n=M$), 
$ {\tilde \Psi}^{  \alpha_1 , \, \alpha_2 , \cdots , \, \alpha_N }_{ 
n_1 , \,  n_2 , \cdots , \, n_N } (x; \gamma ) $ (\ref {b19}) represents
completely symmetric (antisymmetric) eigenfunctions of 
 $SU(M)$ bosonic (fermionic) oscillators.  So, for the case $m, \,n \neq 0$,
we may say that the correlated state vectors 
(\ref {b19}) would represent the eigenfunctions of $N$ number of 
$SU(m|n)$ supersymmetric harmonic oscillators. 

Due to the existence of symmetry condition (\ref {b20}), we find that all 
independent eigenfunctions of $SU(m|n)$ supersymmetric harmonic oscillators 
can be obtained  uniquely through the following occupation
number representation. Since, at present, spins behave as nondynamical degrees
of freedom,  all `single particle states' for this occupation number 
representation may be constructed by taking $m+n$ copies of each energy 
eigenstate associated with a spinless harmonic oscillator $-$
the first $m$ copies being bosonic in nature 
and last $n$ copies being fermionic 
in nature.  As usual, any bosonic single particle state can
be occupied with arbitrary number of particles and 
 any fermionic single particle state can hold at most one particle.
By filling up such bosonic and fermionic single particle states 
 through $N$ number of particles, we can easily identify all independent 
eigenfunctions of the form (\ref {b19}). So, 
this occupation number representation of states 
 (\ref {b19}) gives us a very convenient way of 
 analysing the spectrum of $SU(m|n)$ supersymmetric harmonic 
oscillators.  In particular we may verify that, similar 
to pure bosonic or fermionic case, the spectrum 
  of $SU(m|n)$ supersymmetric harmonic oscillators is also 
equally spaced. But, it is interesting to note further that, the ground state 
energy of $SU(m|n)$ supersymmetric harmonic oscillators coincides with
that of the pure bosonic case, instead of pure fermionic case. 
Since $M$ number of fermionic 
 single particle states with zero energy can hold at most $M$ number of 
 particles,  a nonzero ground state energy is obtained for a pure 
fermionic system when $N>M$. However, if at least one bosonic
 single particle state with zero energy is available, we can fill up
that state through all available particles. Consequently we find that, 
irrespective of the values of $m$ and $n$, the ground state energy of 
 $SU(m|n)$ supersymmetric harmonic oscillators would always be zero.
Moreover, by using the above mentioned occupation number representation, 
 we obtain the degeneracy ($D_g$) of ground state 
 for $SU(m|n)$ supersymmetric harmonic oscillators as
\beq
D_g ~=~ \sum_{k=0}^n \, {  {( N+m-k-1)! \, n! } \over 
          {(N-k)! \, (m-1)! \,  k! \, (n-k)! }  } \, .
\label {b21}
\eeq
It is worth observing that this degeneracy factor 
crucially depends on the values of $m$ and $n$, and reproduces 
the degeneracy of the ground state for $SU(M)$
bosonic oscillators  [8] at the 
special case $m=M,~n=0$. Similarly, one can demonstrate
that the degeneracy of higher energy levels, appearing in the spectrum of  
$SU(m|n)$ supersymmetric harmonic oscillators, would also 
depend on the values of $m$ and $n$.

Due to the relation (\ref {b18}), where ${\cal S}$ acts as a 
nonsingular operator, there exists a one-to-one correspondence between 
the independent eigenfunctions of 
 spin Calogero model (\ref {b5}) and $SU(m|n)$ supersymmetric oscillators.
Consequently, up to a constant energy shift ($E_g$) for all energy levels, 
the spectrum of spin Calogero model (\ref {b5}) exactly coincides 
with the spectrum of $SU(m|n)$ supersymmetric 
harmonic oscillators. However it is well known that,
up to the same constant energy shift for all energy levels, 
the spectrum of spinless Calogero model (\ref {b7}) exactly coincides
with the spectrum of $N$ number of spinless bosonic harmonic oscillators.
As the eigenvalues of both $SU(m|n)$ supersymmetric oscillators 
and spinless bosonic oscillators depend linearly on the coupling
constant $\omega $,  it is clear that eqn.(\ref {b8}) can be used 
even for any finite value of $\omega $ (though the eigenfunctions of 
 spin Calogero model (\ref {b5}) will factorise only at 
the `freezing limit').  Moreover, since the spectra of both 
 $SU(m|n)$ supersymmetric oscillators and spinless bosonic oscillators 
are equally spaced, it automatically follows from eqn.(\ref {b8}) that
the spectrum of $SU(m|n)$ SP model (\ref {a7}) would also be equally spaced 
for any choice of $m$ and $n$.  But it is natural to expect that,
similar to the case of $SU(m|n)$ supersymmetric oscillators,
the degeneracy of energy levels for $SU(m|n)$ SP model (\ref {a7}) 
would crucially depend on the values of $m$ and $n$.  So, it should be
interesting to derive the partition function of SP model 
(\ref {a7}), where all information about these degeneracy 
factors is encoded.

\vspace{1cm}

\noindent \section { Partition function of $SU(m|n)$ SP model }
\renewcommand{\theequation}{3.{\arabic{equation}}}
\setcounter{equation}{0}

\medskip

In the previous section we have observed that, similar to the pure bosonic 
case, the ground state energy of $SU(m|n)$ supersymmetric harmonic 
oscillators would always be zero. So, in analogy with this pure bosonic 
(i.e., ferromagnetic) case [8], we may now put $\omega =1$ in eqn.(\ref {b8}) 
and subsequently use this equation to obtain a relation like 
\beq
Z^{(m|n )}_N (q) ~=~ 
{ {\hat Z}^{(m|n )}_N (q) \over {\hat Z}^{(1|0 )}_N (q) } \, ,
\label {c1}
\eeq
where $q= e^{- {1 \over kT}}$ , 
$Z^{(m|n )}_N (q) $ and 
$ {\hat Z}^{(m|n )}_N (q)$ denote the canonical partition functions of 
 $SU(m|n)$ SP model (\ref {a7}) and 
$SU(m|n)$ supersymmetric harmonic oscillators (where $\omega =1$)
respectively. Due to the above mentioned notations, 
$ {\hat Z}^{(1|0 )}_N (q)$ 
and ${\hat Z}^{(0|1 )}_N (q)$  denote the canonical partition functions 
of $N$ number of spinless bosonic and fermionic harmonic oscillators 
respectively. It is well known that, the partition functions of such 
spinless bosonic and fermionic oscillators are given by [8]
\beq
 {\hat Z}^{(1|0 )}_N (q) ~=~ { 1\over (q)_N } \, , ~~~
 {\hat Z}^{(0|1 )}_N (q) ~=~ q^{ N(N-1) \over 2 } \,
 { 1\over (q)_N } \, , 
\label {c2}
\eeq
where the standard notation: $(q)_N = (1-q) (1-q^2) \cdots (1-q^N) \, $ 
(and $(q)_0 = 1 $) is used.
So, for calculating $Z^{(m|n )}_N (q) $ with the help of eqn.(\ref {c1}), 
we have to find out only the partition function of $N$ number of 
$SU(m|n)$ supersymmetric oscillators. 

To this end, however,
we consider at first the grand canonical partition function of 
$SU(m|n)$ supersymmetric oscillators. Such 
grand canonical partition function may be denoted by 
$ {\hat {\cal Z}}^{(m|n )}(q, y)$, where $y=e^{- \mu }$ and 
$\mu $ is the chemical potential of the system. So, according 
to our notation, $ {\hat {\cal Z}}^{(1|0 )}(q, y)$ and 
$ {\hat {\cal Z}}^{(0|1)}(q, y)$ denote the 
grand canonical partition functions of 
spinless bosonic and fermionic harmonic oscillators respectively.
As usual, the grand canonical partition function 
of $SU(m|n)$ supersymmetric oscillators can be 
related to the corresponding partition function through the following power
series expansion in variable $y$:
\beq
 {\hat {\cal Z}}^{(m|n )}(q, y) ~=~ \sum_{N=0}^{\infty } \, y^N \,
 {\hat Z}^{(m|n )}_N (q) \, 
\label {c3}
\eeq
where it is assumed that $ {\hat Z}^{(m|n )}_0 (q) = 1$.
In the previous section we have found that, all independent
eigenfunctions of $SU(m|n)$ supersymmetric harmonic oscillators can be 
obtained through an occupation number representation, 
where the corresponding single particle states are constructed 
by taking $m+n$ copies of each energy 
eigenstate associated with a spinless harmonic oscillator $-$
the first $m$ copies being bosonic in nature and last $n$ copies being
 fermionic in nature.  By exploiting this result, it is easy to prove that 
the grand canonical partition function of $SU(m|n)$ supersymmetric 
 oscillators can be expressed through those of 
 spinless bosonic and fermionic oscillators as
\beq
{\hat {\cal Z}}^{(m|n )}(q, y) ~=~ 
 \left [ {\hat {\cal Z}}^{(1|0 )}(q, y) \right ]^m 
 \left [ {\hat {\cal Z}}^{(0|1 )}(q, y) \right ]^n \, .
\label {c4}
\eeq
Next, we substitute the power series expansion (\ref {c3}) to the place of
all grand canonical partition functions appearing in the above 
equation. Subsequently, we  compare the coefficients of 
 $y^N$ from both sides of eqn.(\ref {c4}), and readily find that 
 the canonical partition function of $SU(m|n)$ supersymmetric 
harmonic oscillators may also be related to those of 
 spinless bosonic and fermionic oscillators:
\beq
 {\hat Z}^{(m|n )}_N (q)  \,  ~= ~ \,  \sum_{ \sum_{i=1}^m a_i + 
  \sum_{j=1}^n b_j =N } ~
 \prod_{i=1}^m   {\hat Z}^{(1|0 )}_{a_i}(q) \,
 \prod_{j=1}^n   {\hat Z}^{(0|1 )}_{b_j}(q) \, ,
\label {c5}
\eeq
where $a_i$s and $b_j$s are some nonnegative integers.
By substituting the known partition functions (\ref {c2})
of spinless bosonic and fermionic oscillators 
to the above equation, one may now obtain an explicit expression for
 the partition function of $SU(m|n)$ supersymmetric oscillators as
\beq
 {\hat Z}^{(m|n )}_N (q) \, ~=~  \,  \sum_{ \sum_{i=1}^m a_i + 
  \sum_{j=1}^n b_j =N } ~
 {  q^{ \sum_{j=1}^n {b_j (b_j -1 ) \over 2} }  \over 
 (q)_{a_1} (q)_{a_2} \cdots  (q)_{a_m} \cdot (q)_{b_1} (q)_{b_2}
  \cdots  (q)_{b_n}  } \, .
\label {c6}
\eeq

Finally, by using eqns.(\ref {c1}), (\ref {c2}) and (\ref {c6}), we 
 derive the exact canonical partition function of $SU(m|n)$ SP model 
(\ref {a7}) as
\beq
  Z^{(m|n )}_N (q) \, ~=~ \,  \sum_{ \sum_{i=1}^m a_i + 
  \sum_{j=1}^n b_j =N } ~
 {  { (q)_N   \, \cdot \, q^{  \sum_{j=1}^n  {b_j (b_j -1 ) \over 2}  } }
  \over (q)_{a_1} (q)_{a_2} \cdots  (q)_{a_m} \cdot (q)_{b_1} (q)_{b_2}
  \cdots  (q)_{b_n}  } \, .
\label {c7}
\eeq
Since, there exists an upper bound on the highest energy eigenvalue
of SP model (\ref {a7}), the partition function
  $ Z^{(m|n )}_N (q) $ (\ref {c7}) evidently yields some new 
$q$-polynomials which are characterised by the values of $m$ and $n$.
The coefficients of various powers of $q$, appearing in such novel 
$q$-polynomial, would represent the degeneracy factors of corresponding 
energy levels associated with  SP model (\ref {a7}).
It is interesting to note that, by putting $m=M,~n=0$ to the 
 expression (\ref {c7}), one can exactly reproduce the partition 
function of ferromagnetic Polychronakos spin chain [8].
On the other hand, for the limiting case $m=0, ~n=M$, eqn.(\ref {c7}) would
 reproduce the partition function of anti-ferromagnetic Polychronakos
 spin chain up to an insignificant multiplicative factor.
 This is because, in pure fermionic case, a system of harmonic 
oscillators with $M$ spin degrees of freedom gives some nonzero 
ground state energy when $N>M$.
So, the r.h.s. of eqn.(\ref {c1}) must be modified through a multiplicative
factor for taking into account such nonzero ground state energy [8].
Consequently the expression (\ref {c7}) for partition function, 
which we have derived for the supersymmetric case (i.e., when both $m$ and 
$n$ are nonzero), should also be modified by the same 
factor at the pure fermionic limit. 

With the help of a relation: $ \left (q^{-1} \right )_l = 
(-1)^l \, q^{ - \, {l(l+1) \over 2 }} \, (q)_l $, we find that
  the partition function (\ref {c7}) of SP model 
satisfies a remarkable duality condition given by
\beq
  Z^{(m|n )}_N (q) ~=~ q^{ N(N-1) \over 2} \, 
  Z^{(n|m )}_N (q^{-1}) \, ,
\label {c8}
\eeq
where $m$ and $n$ may be chosen as any nonzero integer.
Due to this duality condition, one can write down a relation of the form 
\beq 
  D^{(m|n )}_N (E) ~=~ 
  D^{(n|m )}_N \left ( { N(N-1) \over 2} - E \right ) \, ,
\label {c9}
\eeq
where $D^{(m|n )}_N (E) $ denotes the degeneracy factor 
associated with energy eigenvalue $E$ of $SU(m|n)$ SP model.
Moreover, by using the duality condition (\ref {c8}),
along with the fact that the ground state energy of Hamiltonian 
(\ref {a7}) is zero, it is easy to show that 
\beq 
 E_{\rm max} = { N(N-1) \over 2} \,  
\label {c10}
\eeq
represents the highest energy eigenvalue of the $SU(m|n)$ SP model.
In this context one may notice that,
the highest energy eigenvalue of ferromagnetic or anti-ferromagnetic 
 $SU(M)$ Polychronakos spin chain (\ref {a1}) is given by [8]
\beq
 E_{ \rm max} = { M -1 \over 2M} N^2 - { t (M-t) \over 2M } \, ,
\label {c11}
\eeq
where $ t = N ~{\rm  mod } ~M$.
 Thus we curiously find that, in contrast to the case of 
 $SU(M)$ Polychronakos spin chain (\ref {a1}), the 
highest energy eigenvalue of $SU(m|n)$ SP model (\ref {a7}) does not
depend at all on the values of $m$ or $n$. 

For obtaining some insight about the 
above mentioned difference between the highest energy eigenvalues 
(\ref {c10}) and (\ref {c11}), we finally consider the 
`motif' picture, which was used to analyse the degeneracy of eigenfunctions 
for $SU(M)$ HS as well as Polychronakos spin chain through the corresponding 
symmetry algebra [11,13]. As is well known, motifs of these spin chains 
are made of binary digits like `0' and `1'. Moreover, for a spin chain
with $N$ number of lattice sites, these binary digits would form motifs 
of length $N-1$. So we can write a motif as 
$(a_1 a_2 \cdots a_{N-1})$, where $a_i \in [0,1]$ and each motif
 represents a class of degenerate eigenfunctions which yield 
an irreducible representation of $Y(gl_M)$ Yangian algebra [11,18].
For the case of $SU(M)$ Polychronakos spin chain, the energy eigenvalue 
corresponding to $(a_1 a_2 \cdots a_{N-1})$ motif is given by [13]
\beq
 E_{ (a_1 a_2 \cdots a_{N-1}) } ~=~ \sum_{r=1}^{N-1} ~  r \, a_r  \, .
\label {c12}
\eeq
However, for the case of $SU(M)$ Polychronakos and HS spin chain, 
there exists a `selection rule' which forbids the occurrence of $M$ number of 
consecutive `1's in any motif. By combining this selection rule 
along with eqn.(\ref {c12}), one can derive the highest energy 
eigenvalue (\ref {c11}) of $SU(M)$ Polychronakos model and also understand 
why this eigenvalue crucially depends on the value of $M$. It is worth noting 
that, the above mentioned motif picture can also be used to analyse the 
spectrum of $SU(m|n)$ supersymmetric HS spin chain [18,28]. But, for this
supersymmetric case,  there exists no `selection rule' and 
 binary digits like `0' and `1' can be chosen freely for constructing 
a motif of length $N-1$. Being motivated by the case of 
 supersymmetric HS spin chain, we may now conjecture that the motifs 
corresponding to SP model (\ref {a7}) are 
 also free from any `selection rule' and eqn.(\ref {c12})
again gives the corresponding energy eigenvalues. 
So, the spectrum of $SU(m|n)$ SP model contains $2^{N-1}$ 
number of motifs, which can be constructed 
by filling up $N-1$ positions with `0' or `1' in all possible ways.
By using eqn.(\ref {c12}), it is easy to 
check that the motif $(11 \cdots 1)$  yields the highest energy eigenvalue 
(\ref {c10}). Thus the absence of any `selection rule', 
for the motifs corresponding to SP model, turns out to be the main reason
behind the remarkably simple expression (\ref {c10}).

\vspace{1cm}

\noindent \section { Concluding Remarks }
\renewcommand{\theequation}{4.{\arabic{equation}}}
\setcounter{equation}{0}

In this paper we have investigated the spectrum as well as
 partition function of $SU(m|n)$ supersymmetric Polychronakos (SP) model 
(\ref {a7}) and the related spin Calogero model (\ref {b5}).
The similarity transformation (\ref {b11}), which maps
the spinless Calogero model of distinguishable particles to 
decoupled oscillators, and the projection operator 
(\ref {b16}) have played a key role in our derivation for the
spectrum of spin Calogero model (\ref {b5}). Thus we have found that,
up to a constant energy shift for all states, the spectrum of this
spin Calogero model is exactly the same with the spectrum of 
decoupled $SU(m|n)$ supersymmetric harmonic oscillators. Furthermore,
by using the occupation number representation associated with 
 $SU(m|n)$ supersymmetric harmonic oscillators, we have obtained 
the exact partition function for this spin Calogero model.

It turned out that, the above mentioned spin Calogero model 
reproduces the SP model (\ref {a7}) at the `freezing limit'. 
Consequently, by factoring out the contributions due to
 dynamical degrees of freedom from the spectrum as well as partition 
function of spin Calogero model (\ref {b5}), one can compute the spectrum and 
partition function for $SU(m|n)$ SP model. By following this procedure
we have found that, similar to the 
non-supersymmetric case, the spectrum of $SU(m|n)$ SP model is also equally 
spaced.  However, the degeneracy factors of corresponding energy levels 
crucially depend on the values of $m$ and 
$n$. As a result, we get some novel $q$-polynomials which represent the 
partition functions of SP models.
Moreover, by interchanging the bosonic and fermionic degrees of freedom, 
we obtain a duality relation among the partition functions of SP models.

As a future study, it should be interesting to find out the 
Lax operators and conserved quantities for the 
$SU(m|n)$ SP model (\ref {a7}). Moreover, in parallel to the case
of $SU(m|n)$ Haldane-Shastry model [18], one might be able to show that 
the $SU(m|n)$ SP model also exhibits the $Y(gl_{(m|n)})$ 
super-Yangian symmetry. Such 
super-Yangian symmetry of SP model may turn out to be very helpful in 
analysing the degeneracy patterns for its spectrum. However, 
information about these degeneracy patterns is also encoded in our 
 expression of partition function (\ref {c7}). So, there should exist an 
intriguing connection between the partition function (\ref {c7}) and the 
motif representations for super-Yangian algebra. In particular,
the partition function (\ref {c7}) might lead to a 
supersymmetric generalisation of well-known 
 Rogers-Szeg${\ddot {\rm o}}$ (RS) polynomial. The recursion 
relation among these supersymmetric
 RS polynomials may then be used to find out the motifs representations 
 and degenerate multiplets associated with super-Yangian algebra.
We hope to report about such supersymmetric RS polynomials and 
related motifs in a forthcoming publication [34].

\bigskip

\noindent {\bf Acknowledgments }

We like to thank K. Hikami for many illuminating discussions.
One of the authors (BBM) likes to acknowledge  Japan Society 
for the Promotion of Science for a fellowship (JSPS-P97047) which 
supported this work.

\newpage 
\leftline {\large \bf References } 
\medskip 
\begin{enumerate}

\item F. Calogero: J. Math. Phys. {\bf 10} (1969) 2191.

\item B. Sutherland: Phys. Rev. A {\bf 5} (1972) 1372.

\item F.D.M. Haldane: Phys. Rev. Lett. {\bf 60} (1988) 635.

\item B.S. Shastry: Phys. Rev. Lett. {\bf 60} (1988) 639.

\item M.A. Olshanetsky and A.M. Perelomov: Phys. Rep. {\bf 94} (1983) 313.

\item A.P. Polychronakos: Phys. Rev. lett. {\bf 70} (1993) 2329.

\item H. Frahm: J. Phys. A {\bf 26} (1993) L473.

\item A.P. Polychronakos: Nucl. Phys. B {\bf 419} (1994) 553.

\item H. Ujino, K. Hikami and M. Wadati: J. Phys. Soc. Jpn. {\bf 61}
 (1992) 3425.

\item K. Hikami and M. Wadati: J. Phys. Soc. Jpn. {\bf 62} (1993) 4203;
        Phys. Rev. Lett. {\bf 73} (1994) 1191.

\item F.D.M. Haldane, Z.N.C. Ha, J.C. Talstra, D. Bernard and V. Pasquier:
          Phys. Rev. Lett. {\bf 69} (1992) 2021.

\item D. Bernard, M. Gaudin, F.D.M. Haldane and V. Pasquier:
          J. Phys. A {\bf 26} (1993) 5219.

\item K. Hikami: Nucl. Phys. B {\bf 441} [FS] (1995) 530. 

\item Z.N.C. Ha: Phys. Rev. Lett. {\bf 73} (1994) 1574;
          Nucl. Phys. B {\bf 435} [FS] (1995) 604.

\item M.V.N. Murthy and R. Shankar: Phys. Rev. Lett. {\bf 73} (1994) 3331.

\item F. Lesage, V. Pasquier and D. Serban: Nucl. Phys. B {\bf 435}
 [FS] (1995) 585. 

\item B. Sutherland and B.S. Shastry: Phys. Rev. Lett. {\bf 71} (1993) 5.

\item F.D.M. Haldane:  Proc. 16th Taniguchi Symp., Kashikojima,
       Japan, (1993) eds. A. Okiji and N. Kawakami (Springer, Berlin,
       1994).

\item B.D. Simons, P.A. Lee and B.L. Altshuler:  Nucl. Phys. B {\bf 409}
          (1993) 487.

\item A.P. Polychronakos: {\it Generalised statistics in one dimension },
      hep-th/9902157 (Les Houches Lectures, Summer 1998)

\item K. Hikami: J. Phys. A {\bf 28} (1995) L131.

\item K. Hikami: J. Phys. Soc. Jpn. {\bf 64} (1995) 1047.

\item G.E. Andrews: {\it The Theory of Partitions } 
      (Addison-Wesley, London, 1976).

\item E. Melzer: Lett. Math. Phys. {\bf 31} (1994) 233.

\item B. Basu-Mallick: Nucl. Phys. B {\bf 482} [FS] (1996) 713.

\item B. Basu-Mallick and A. Kundu: Nucl. Phys. B {\bf 509} [FS] (1998) 705.

\item B. Basu-Mallick: J. Phys. Soc. Jpn. {\bf 67} (1998) 2227.

\item B. Basu-Mallick: Nucl. Phys. B {\bf 540} [FS] (1999) 679.

\item K. Sogo: J. Phys. Soc. Jpn. {\bf 65} (1996) 3097.

\item T. H. Baker and P.J. Forrester: Nucl. Phys. B {\bf 492} (1997) 682.

\item N. Gurappa and P.K. Panigrahi: {\it Equivalence of the 
        Calogero-Sutherland model to free harmonic oscillators},
        cond-mat/9710035.

\item H. Ujino, A. Nishino and M. Wadati: J. Phys. Soc. Jpn. 
      {\bf 67} (1998) 2658.

\item H. Ujino, A. Nishino and M. Wadati: Phys. Lett. A {\bf 249}
       (1998) 459.

\item K. Hikami and B. Basu-Mallick: {\it Supersymmetric Polychronakos 
spin chain. Motif, distribution function and character }
(University of Tokyo Preprint, Under Preparation)

\end{enumerate} 
\end{document}